\pdfoutput=1
\RequirePackage{ifpdf}
\ifpdf % We~are running pdfTeX in pdf mode
\documentclass[pdftex]{sigma}
\else
\documentclass{sigma}
\fi

\usepackage{mathrsfs}
\newcommand{\Res}{\mathop{\rm Res}}

\numberwithin{equation}{section}

\newtheorem*{Theorem*}{Theorem}

 { \theoremstyle{definition}

 }

\begin{document}

\allowdisplaybreaks

\newcommand{\arXivNumber}{2307.02080}

\renewcommand{\PaperNumber}{028}

\FirstPageHeading

\ShortArticleName{Resurgent Structure of the Topological String and the First Painlev\'e Equation}

\ArticleName{Resurgent Structure of the Topological String\\ and the First Painlev\'e Equation}

\Author{Kohei IWAKI~$^{\rm a}$ and Marcos MARI\~NO~$^{\rm b}$}

\AuthorNameForHeading{K.~Iwaki and M.~Mari\~no}

\Address{$^{\rm a)}$~Graduate School of Mathematical Sciences, The University of Tokyo,\\
\hphantom{$^{\rm a)}$}~3-8-1 Komaba, Meguro-ku, Tokyo, 153-8914, Japan}
\EmailD{\href{mailto:k-iwaki@g.ecc.u-tokyo.ac.jp}{k-iwaki@g.ecc.u-tokyo.ac.jp}}

\Address{$^{\rm b)}$~Section de Math\'ematiques et D\'epartement de Physique Th\'eorique, Universit\'e de Gen\`eve,\\
\hphantom{$^{\rm b)}$}~1211 Gen\`eve 4, Switzerland}
\EmailD{\href{mailto:marcos.marino@unige.ch}{marcos.marino@unige.ch}}
\URLaddressD{\url{http://www.marcosmarino.net}}

\ArticleDates{Received September 13, 2023, in final form March 19, 2024; Published online April 02, 2024}

\Abstract{We present an explicit formula for the Stokes automorphism acting on the topological string partition function. When written in terms of the dual partition function, our formula implies that flat coordinates in topological string theory transform as quantum periods, and according to the Delabaere--Dillinger--Pham formula. We first show how the formula follows from the non-linear Stokes phenomenon of the Painlev\'e~I equation, together with the connection between its $\tau$-function and topological strings on elliptic curves. Then, we show that this formula is also a consequence of a recent conjecture on the resurgent structure of the topological string, based on the holomorphic anomaly equations, and it is in fact valid for arbitrary Calabi--Yau threefolds.}

\Keywords{resurgence; topological string theory; first Painlev\'e equation; Borel resummation; Stokes automorphisms}

\Classification{81T45; 14N35; 34M40; 34M55}

\section{Introduction and conclusions}
\label{sec.intro}

Topological string theory has been at the center of many developments in modern mathematical physics.
For example, through string dualities \cite{bkmp,dv} and the topological recursion of \cite{EO07} it is related to
matrix models and the $1/N$ expansion. It has been also shown that certain ``dual'' partition functions appearing in topological string theory are closely related to $\tau$-functions of Painlev\'e equations \cite{EGF, EGFMO,Iwaki19, MO19-2}.
These results have a close relation with the pioneering works~\cite{BLMST-16,GIL-12}.

The basic quantities in topological string theory are given by formal perturbative series in the string coupling constant $g_s$, which
are factorially divergent. One can then ask, in the spirit of the theory of resurgence \cite{ecalle,msauzin},
what is the resurgent structure of these series, i.e., what are the exponentially small
trans-series associated to their Borel singularities, and the associated Stokes constants. This question has been recently addressed
in various works. Based on the trans-series approach of \cite{cesv2, cesv1, cms}, a conjecture for the resurgent structure of topological
string theory on arbitrary Calabi--Yau (CY) threefolds was put forward in \cite{gkkm,gm-multi}. Although topological strings are closely
related to quantum periods (also known as Voros symbols),
their resurgent structure is quite different. For example, it is expected
that the Stokes automorphisms of quantum periods are given by the
Delabaere--Dillinger--Pham (DDP) formula \cite{DDP93,ddpham,DP99} (see \cite{dml, gm-qp} for recent
arguments in that direction), but the Stokes automorphisms act on the topological string partition function with a more complicated
structure, and no closed formula is known for them.

In this paper we provide such a closed formula. It reads
\begin{equation} \label{eq:main-formula}
%\boxed{
{\mathfrak S} Z(\nu; g_s) = \exp\bigl\{ a \operatorname{Li}_2 \bigl( {\rm e}^{-g_s \partial} \bigr)-
g_s a \log \bigl(1- {\rm e}^{-g_s \partial} \bigr) \partial \bigr\} Z(\nu; g_s),
%}
\end{equation}
where $a$ is a Stokes constant, $\nu$ is the flat coordinate (or period) parametrizing the moduli space, and $\partial$ is the derivative with reference to $\nu$.\footnote{We have written the formula in a somewhat schematic way, and for a simple case, involving a single modulus; additional details, generalizations and clarifications can be found below.}

Although the partition function transforms in a relatively complicated way under the Stokes automorphism, the dual partition function (obtained from the original one after a discrete Fourier transform)
transforms in a very simple way:
\begin{equation}
\label{eq:main-formula-2}
%\boxed{
{\mathfrak S} \tau(\nu,\rho; g_s) = {\rm e}^{a \operatorname{Li}_2({\rm e}^{2 \pi {\rm i} \rho/ g_s})}
\tau\big(\nu -a g_s \log\big(1 - {\rm e}^{2 \pi {\rm i} \rho/ g_s}\big) , \rho; g_s \big),
%}
\end{equation}
i.e., it picks up a prefactor involving the dilogarithm function, and its argument transforms
indeed as a quantum period, following the DDP formula!

In this paper, we give two lines of argument that lead to \eqref{eq:main-formula} and \eqref{eq:main-formula-2}.
The first one is based on the connection found in \cite{Iwaki19}
between $\tau$-functions of the first Painlev\'e equation, and a dual partition function in
topological string theory. By using this connection, one can show that the non-linear Stokes
phenomenon of the first Painlev\'e equation leads to the above transformation formula for the
dual partition function under a Stokes automorphism.
Although this derivation is based on a particular example of the topological string partition function,
the resulting formula turns out to be universally valid.
To show this, we develop a second argument based on the conjectures
for the resurgent structure of the topological string proposed in \cite{gkkm,gm-multi},
and we provide a derivation of \eqref{eq:main-formula} in the spirit of alien calculus.
Since the conjectures of \cite{gkkm,gm-multi} apply to any topological string partition function which satisfies the holomorphic anomaly equations (HAE) of \cite{bcov}, we conclude that \eqref{eq:main-formula} applies universally to closed topological strings on arbitrary backgrounds. It can be extended to refined and real topological strings by using similar ideas and techniques, see \cite{amp,ms}.

As is mentioned above, the derivation based on \cite{Iwaki19} naturally relates
the DDP formula to our main formula.
The DDP formula has a close relationship with the {BPS invariants}
in class~${\mathcal S}$ theories,
which are defined by a weighted counting of saddle connections in the spectral network~\cite{GMN08}.
In fact, the DDP formula for the quantum period $V_\mu$ reads
\begin{equation}
{\rm e}^{V_\mu} \mapsto
{\rm e}^{V_\mu} \bigl(1 + \sigma(\gamma) {\rm e}^{V_\gamma}\bigr)^{\Omega(\gamma) \langle \mu, \gamma \rangle},
\end{equation}
where $\gamma$ is the cycle on the spectral curve associated with the saddle connection,
the sign $\sigma(\gamma) \in \{\pm 1 \}$ is specified by the type of saddle connection,
and $\Omega(\gamma)$ is the BPS invariant; see, e.g., \cite{Br-Sm,GMN08}.
In view of this, we expect the Stokes constant $a$ appearing in
\eqref{eq:main-formula}, \eqref{eq:main-formula-2} and in the formulas in Section~\ref{sec-HAEderivation} to be given by a BPS invariant $\Omega(\gamma)$. The precise relation is
\begin{equation}
\label{aomega-intro}
a= \frac{\Omega(\gamma)}{2 \pi {\rm i}}.
\end{equation}
In the more general
case considered in this paper, the $\Omega(\gamma)$ are given by the Donaldson--Thomas invariants of the CY threefold.
Here, the cycle $\gamma$ is dual to the Borel singularity ${\mathcal{A}}$ underlying the
Stokes discontinuity, and as we explain below ${\mathcal{A}}$ is an integral period of the CY geometry (up to an overall normalization).
This identification between Stokes constants appearing in the resurgent structure of the topological string and BPS invariants was
already suggested in \cite{gkkm,gm-multi}, and it is also consistent with the results of \cite{IK-I, IK-II}.
Further evidence for this identification appears in the recent paper \cite{gu-res}.

We should mention that close cousins of \eqref{eq:main-formula-2} have appeared before in different, but related contexts.
In \cite{AP}, the transformation properties of a certain family of theta series defined on the (twistor space of the) hypermultiplet
moduli space of CY threefolds under wall-crossing were studied. In the special case $k=1$, the theta series
(1.5) in \cite{AP} reduces to the dual topological partition function \cite{APP}, and their wall-crossing formula (1.9)
agrees with our formula for the Stokes automorphism. In addition, for the formula to agree, one must
have the relation \eqref{aomega-intro}, giving in this way additional evidence for the identification between Stokes
constants and Donaldson--Thomas invariants. A detailed comparison between
the wall-crossing formula of \cite{AP} and our formula will be made in Section \ref{sec-HAEderivation}.
A similar wall-crossing formula was also found in \cite{CLT}
for the dual partition functions of some ${\mathcal N}=2$ gauge theories. Let us point out that in
\cite{APP, AP, CLT} the transformation of the
flat coordinate according to the DDP formula is essentially built in,
while in our case it follows in a more indirect (and surprising) way from the resurgent structure of the topological string perturbative series.
It would be very interesting to understand better the relation between
\cite{APP,AP, CLT} and our approach.

The result obtained in this paper can be understood as relating the Stokes automorphism acting on the
topological string partition function, to the Stokes automorphism acting on quantum periods.
It might have a connection to the blowup formula which relates in a similar way the topological
string free energy to the Nekrasov--Shatashvili free energy \cite{gg,jeong-n}. This idea has also been developed
in \cite{gu-res} and it might lead to a different derivation of our main formula. It would be also
interesting to study the relation between our work and a series of papers by Bridgeland
\cite{Bri-23, Bri-19, BriM}.

 After this paper was finished, we were informed by R.~Schiappa and M.~Schwick that in forthcoming work %~\cite{sst}
 they address similar issues and obtain related results, albeit with different methods.

\section[From non-linear Stokes phenomenon of the first Painlev\'e transcendents to the main formula]{From non-linear Stokes phenomenon of the first\\ Painlev\'e transcendents to the main formula}
\label{painleve}

In this section, we give a derivation of the formula
\eqref{eq:main-formula-2}
for the topological recursion partition function
through the analysis of the non-linear Stokes phenomenon of {the first} Painlev\'e equation.
Our derivation is based on the exact WKB theoretic approach to
the Stokes multipliers of the isomonodromy system
associated with {the first} Painlev\'e equation,
which was developed in \cite[Section~5]{Iwaki19}.
We note that our result is closely related to \cite{bssv,Takei1998, vv},
and we will make a comment in the end of this section.
We also note that our derivation is based on several conjectures
on the Borel summability of partition function and wave function etc.,
which are naturally expected from the viewpoint of the exact WKB analysis
of Schr\"odinger equations (see \cite{KT05}, for example).
See \cite{ecalle, msauzin} for materials in Borel summation method and resurgence.

\subsection{Topological recursion and Painlev\'e I}

Let us briefly review the relationship between topological recursion and the first Painlev\'e equation.

We focus on the topological recursion partition function
\begin{equation} \label{eq:Z-TR}
Z(t,\nu; g_s) = \exp\biggl( \sum_{g \ge 0} g_s^{2g-2} F_g(t,\nu) \biggr)
\end{equation}
defined from a family of genus 1 spectral curves of the form
\begin{equation} \label{eq:spcurve}
\Sigma\colon\
y^2 = 4x^3 + 2t x + u(t,\nu)
\end{equation}
parametrized by independent complex parameters $t$ and $\nu$.
Here, $u(t,\nu)$ is a locally-defined function given by the implicit relation
\begin{gather}
\nu = \frac{1}{2 \pi {\rm i}} \oint_A y\, {\rm d}x
\end{gather}
after fixing a choice of a symplectic basis $A$, $B$ of $H_1(\Sigma, {\mathbb Z})$.
See Appendix \ref{appendix:TR} for the definition of $F_g$
(see also \cite{EO07, Iwaki19}).
These quantities depend on the choice of the symplectic basis,
and we will choose a specific one to derive our main formula in Section~\ref{subsection:DDP}.

It was shown in \cite{Iwaki19} that the discrete Fourier transform
\begin{equation} \label{eq:tau-PI}
\tau(t,\nu,\rho; g_s) = \sum_{k \in {\mathbb Z}} {\rm e}^{2 \pi {\rm i} k \rho/g_s}
Z(t, \nu+k g_s ; g_s)
\end{equation}
gives a $\tau$-function for the {Painlev\'e I equation}\footnote{This construction is now generalized to all Painlev\'e equations;
see \cite{EGF, EGFMO, MO19-2}.}
\begin{equation} \label{eq:PI}
g_s^2 \frac{{\rm d}^2 q}{{\rm d}t^2} = 6q^2 + t.
\end{equation}
That is,
\begin{equation} \label{eq:q-and-tau}
q(t,\nu,\rho; g_s) = - g_s^2 \frac{{\rm d}^2}{{\rm d}t^2} \log \tau(t,\nu,\rho; g_s)
\end{equation}
is a formal solution of \eqref{eq:PI}.
Actually, the formal series \eqref{eq:tau-PI} and \eqref{eq:q-and-tau}
(and \eqref{eq:TR-wave-function} below)
are not a usual power series in $\hbar$, but can be regarded as a two-sided trans-series
(which contain both positive and negative exponential factors).
These exponential terms can be summed up to $\vartheta$-functions.
The discrete Fourier transform can be regarded as a non-perturbative
correction to perturbative partition function; see \cite{EGFMO, emo, Iwaki19} for details.
The parameters $\nu$ and $\rho$ are regarded as integration constants
parametrizing the general solution of Painlev\'e I.
The formula \eqref{eq:tau-PI} is analogous to the formula
obtained in \cite{BLMST-16, GIL-12}.

To derive our main formula \eqref{eq:main-formula},
we will also use the isomonodromy system associated with Painlev\'e I \cite{JMU}:
\begin{gather}\label{eq:Lax-PI}
\begin{split}
&\left[ g_s^2 \frac{\partial^2}{\partial x^2} -
\frac{g_s}{x-q} \left( g_s \frac{\partial}{\partial x} - p \right)
- \left(4x^3 + 2t x + 2 H\right) \right] Y = 0,
\\
&\left[ g_s \frac{\partial}{\partial t} - \frac{1}{2(x-q)}
\left( g_s \frac{\partial}{\partial x} - p \right)
\right] Y = 0,
\end{split}
\end{gather}
where
\begin{equation}
p = g_s \frac{{\rm d}q}{{\rm d}t}, \qquad
H = \frac{1}{2}p^2 - 2q^3 - t q.
\end{equation}
It is well known that the compatibility condition of the system \eqref{eq:Lax-PI}
of PDEs is given by the Painlev\'e I equation.
The compatibility implies that the {Stokes multipliers}\footnote{This paper discusses {two} types of Stokes phenomenon.
The first one is related to the limit $g_s \to 0$,
while the second one is related to $x \to \infty$.
Our main formulas \eqref{eq:main-formula} and \eqref{eq:main-formula-2}
are related to the first type Stokes phenomenon,
but we analyze the second type as well to derive the main formulas.}
of the first equation of
the system \eqref{eq:Lax-PI}, which is a linear ODE with an irregular singular point $x=\infty$
of Poincar\'e rank $5/2$, are independent of the isomonodromic time $t$.
It was also shown in \cite{Iwaki19} that
\begin{equation} \label{eq:TR-wave-function}
Y_{\pm}(x,t,\nu,\rho; g_s)
= \frac{\sum_{k \in {\mathbb Z}} {\rm e}^{2 \pi {\rm i} k \rho / g_s}
 Z(t, \nu + k g_s; g_s) \, \chi_{\pm}(x,t,\nu+k g_s; g_s)}
{\sum_{k \in {\mathbb Z}} {\rm e}^{2 \pi {\rm i} k \rho / g_s}
Z(t, \nu + k g_s; g_s) }
\end{equation}
gives a formal solution for the isomonodromy system.
Here, $\chi_{\pm}$ is a WKB-type formal series
defined by the ``{quantum curve formula}'' as follows\footnote{Precisely speaking, we must regularize the integral in \eqref{eq:WKB-series}
for $(g,n) = (0,1)$ and $(0,2)$; see \cite{Iwaki19}.}
(see \cite{BE16}, for example):
\begin{gather}
\chi_{\pm}(x,t,\nu; g_s)
\label{eq:WKB-series}\\
= \exp\Biggl(
\sum_{\substack{g \ge 0 \\ n \ge 1}} \frac{(\pm g_s)^{2g-2+n}}{n!}
\int^{z(x)}_{0} \cdots  \int^{z(x)}_{0}
\left( \omega_{g,n}(z_1, \dots, z_n) - \delta_{g,0} \delta_{n,2}
\frac{{\rm d}x(z_1) {\rm d}x(z_2)}{(x(z_1) - x(z_2))^2} \right)
\Biggr).\notag
\end{gather}
Here, $\omega_{g,n}$'s are the topological recursion correlators
(see Appendix \ref{appendix:TR} for the definition).

The most important formula to have a formal solution to the
isomonodromy system \eqref{eq:Lax-PI} is the following one, which describes
the term-wise (in $g_s$)
analytic continuation along $A$-cycle and $B$-cycle with respect to $x$-variable:
\begin{equation} \label{eq:Voros-symbols}
Y_{\pm}(x,t,\nu,\rho; g_s)
 \mapsto
\begin{cases}
{\rm e}^{\pm 2 \pi {\rm i} \nu/g_s} \, Y_{\pm}(x,t,\nu,\rho;g_s) &
\text{along $A$-cycle}, \\[+.5em]
{\rm e}^{\mp 2 \pi {\rm i} \rho/g_s} \, Y_{\pm}(x,t,\nu,\rho;g_s) &
\text{along $B$-cycle}.
\end{cases}
\end{equation}
See \cite[Theorems 3.9 and 4.8]{Iwaki19} for the derivation of \eqref{eq:Voros-symbols}.
In the spirit of the {exact WKB analysis},
and the {abelianization} in the sense of Gaiotto--Moore--Neitzke \cite{GMN12},
the monodromy and Stokes data of a Schr\"odinger-type linear ODE should be written
in terms of the {quantum periods} on the spectral curve (see \cite[Section~3]{KT05} for more details).
Exponentials of those periods are called {Voros symbols},
which play an important role in the exact WKB analysis (see~\cite{DDP93, KT05, Voros83}, for example).
In view of the property \eqref{eq:Voros-symbols},
it is natural to define the {Voros symbols of the isomonodromy system}
\eqref{eq:Lax-PI} along $A$-cycle and $B$-cycle
by ${\rm e}^{2 \pi {\rm i} \nu/g_s}$ and ${\rm e}^{2 \pi {\rm i} \rho/g_s}$, respectively,
even though the parameter $\rho$ is not an actual period integral on the spectral curve.
These Voros symbols are $t$-independent, so it is natural to expect
that $Y_{\pm}$ satisfies an isomonodromy system.
This is a philosophical remark on the result of \cite{Iwaki19}.

The discrete Fourier transform is absolutely crucial to obtain
the constant monodromy property \eqref{eq:Voros-symbols} and
the quantum curve (i.e., the first equation in \eqref{eq:Lax-PI}) as an ODE.
The property is proved for more general spectral curves in \cite{EGFMO}.
It would be interesting to apply the following method to other Painlev\'e equations.

\subsection{Stokes graph of the isomonodromy system}\label{subsec:Stokes-graph}

Here we also review the discussion of \cite[Section~5]{Iwaki19} on heuristic computation of
the Stokes multipliers of the isomonodromy system.

Let us take the meromorphic quadratic differential
\begin{equation}% \label{eq:quad-diff}
\varphi(x) = \big(4x^3 + 2 t x + u(t,\nu)\big) {\rm d}x^2,
\end{equation}
and define the {Stokes graph} of (the first equation of) the isomonodromy system \eqref{eq:Lax-PI}
as the graph on ${\mathbb P}^1$
as follows:
\begin{itemize}\itemsep=0pt
\item The vertices of the Stokes graph consists of zeros and poles of $\varphi$.
\item The edges of the Stokes graph, called {Stokes curves},
are trajectories of $\varphi$ emanating from a zero of $\varphi$.
\end{itemize}
Here, a (horizontal) {trajectory} of $\varphi$ is any maximal leaf
of the foliation on ${\mathbb P}^1$ defined by{\samepage
\begin{gather}
{\rm Im} \int^{x} \sqrt{\varphi(x)} = {\rm constant}.
\end{gather}
See \cite{Strebel} for more details on trajectories of quadratic differentials.}

\begin{figure}[t]\centering
 \includegraphics[width=43mm]{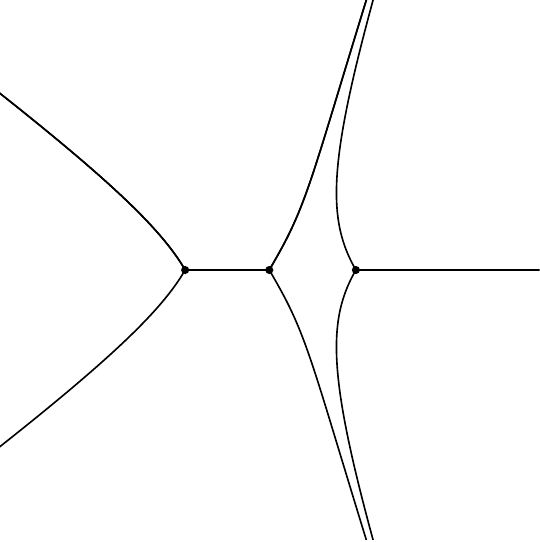} \\
 {$t = t_c$}
\\
 \begin{minipage}{0.35\hsize}
\centering
 \includegraphics[width=43mm]{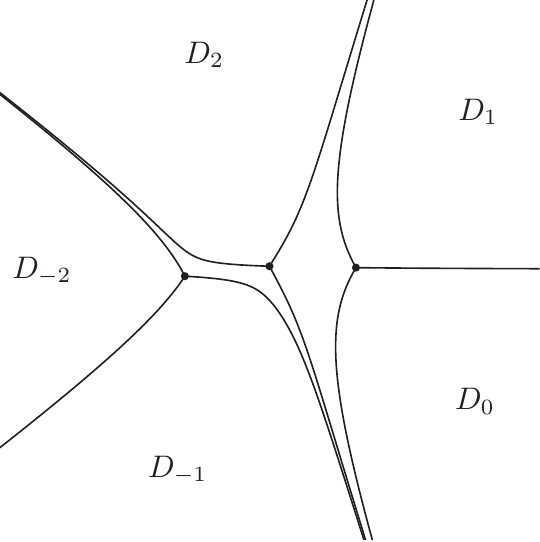} \\
 {$t = t_c - {\rm i} \epsilon$}
 \end{minipage}
 \begin{minipage}{0.35\hsize}
\centering
 \includegraphics[width=43mm]{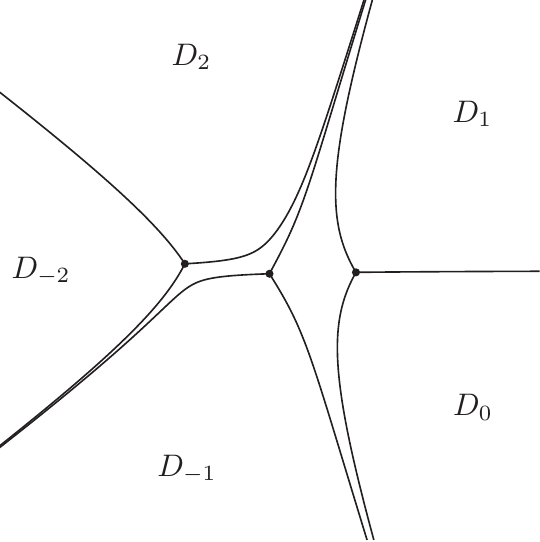} \\
 {$t = t_c + {\rm i} \epsilon$}
 \end{minipage}
 \caption{Stokes graphs for several $t$ (with $\nu = 1$).
 Here we choose $t_c = -5$ and $\epsilon = 1/2$.} \label{fig:mutation}
\end{figure}

The Stokes graph is also known as an example of spectral networks of \cite{GMN12}.
Note that the Stokes graph depends on the parameters $t$ and $\nu$.
Away from the zero locus of the discriminant, $\varphi$ has three simple zeros,
and three Stokes curves emanate from each simple zero.
Each face of the Stokes graph is called a {Stokes region}.
Some examples are shown in Figure \ref{fig:mutation}.
The neighborhood of the irregular singular point $x = \infty$
is divided into five sectors by asymptotic directions of Stokes curves.
The five directions are called {singular directions}.

The main conjectural ansatz for the computation of the Stokes multipliers
of the isomonodromy system \eqref{eq:Lax-PI} around $x=\infty$ are the following:
\begin{itemize}\itemsep=0pt
\item[{\rm (i)}]
If the Stokes graph does not contain any saddle connection
(i.e., a Stokes curve connecting zeros of $\varphi$),
then the partition function \eqref{eq:Z-TR} is Borel summable.
Moreover, the discrete Fourier series
\begin{gather}
{\mathscr T}(t,\nu,\rho; g_s) =
\sum_{k \in {\mathbb Z}} {\rm e}^{2 \pi {\rm i} k \rho/g_s} {\mathcal Z}(t,\nu+k g_s; g_s)
\end{gather}
converges and gives an {analytic $\tau$-function} for Painlev\'e I.
Here, ${\mathcal Z}$ is the Borel sum of the partition function $Z$.

\item[{\rm (ii)}]
Under the same saddle-free condition, the WKB-type series $\chi_\pm$ defined in \eqref{eq:WKB-series}
is Borel summable on each Stokes region. Moreover,
\begin{gather}%\label{eq:Borel-resumed-NPWKB}
{\mathscr Y}_{\pm}(x,t,\nu,\rho;g_s)
= \frac{\sum_{k \in {\mathbb Z}} {\rm e}^{2 \pi {\rm i} k \rho / g_s}
 {\mathcal Z}(t, \nu + k g_s; g_s) \, {\mathcal X}_{\pm}(x,t,\nu+k g_s; g_s)}
{\sum_{k \in {\mathbb Z}} {\rm e}^{2 \pi {\rm i} k \rho / g_s}
{\mathcal Z}(t, \nu + k g_s; g_s) }
\end{gather}
converges and give an analytic solution of the isomonodromy system \eqref{eq:Lax-PI} on the Stokes region.
Here, ${\mathcal X}_{\pm}$ is the Borel sum of $\chi_\pm$.

\item[{\rm (iii)}]
Under the same saddle-free condition, the Borel sums
${\mathcal X}_{\pm}$ of the WKB-type series \eqref{eq:WKB-series}
defined on adjacent faces of the Stokes graph are related by
the {Voros connection formula} \cite{KT05,silverstone,Voros83}
(or the {path-lifting rule} in the sense of Gaiotto--Moore--Neitzke \cite{GMN12}).
\end{itemize}

The assumption (i) is consistent with the conjecture in \cite{dmp-np, gkkm, gm-multi}
which claims the Borel singularities appear on the lattice generated by the integral periods of $y {\rm d}x$.
For a class of genus~$0$ spectral curves, the conjecture is proved rigorously in \cite{IK-I, IK-II}.
Note that the saddle-free condition in (i) is satisfied if all the integral periods of
$y {\rm d}x$ on the spectral curve have a non-zero imaginary part.

The assumption (ii) is inferred from the known result on
the Borel summability of WKB solution of Schr\"odinger equations
(cf.\ \cite{DLS, Nemes, Nikolaev} and an unpublished work by Koike--Sch\"afke).
These known rigorous results are not applicable to our case since $\chi_\pm$
defined in \eqref{eq:WKB-series} does not satisfy an ODE but a PDE,
but we expect that the same statement is true since the PDE
also has~$\Sigma$ as the classical limit (see \cite[Theorem~3.7]{Iwaki19}).
Under the assumption~(ii), we have five canonical solutions \smash{${\mathscr Y}^{(j)}_{\pm}$}
defined in the Stokes region $D_j$ in Figure \ref{fig:mutation} ($j = 0, \pm1, \pm2$ mod $5$).
Then, we define the Stokes multiplier $s_j$ attached to the $j$-th singular direction
$\arg x = 2\pi j/5$ as the non-trivial entry of the Stokes matrix ${\mathcal S}_j$ defined by
\begin{gather}
\big({\mathscr Y}^{(j+1)}_{+}, {\mathscr Y}^{(j+1)}_{-}\big) =
\big({\mathscr Y}^{(j)}_{+}, {\mathscr Y}^{(j)}_{-}\big) \cdot {\mathcal S}_j,
\end{gather}
where
\begin{equation}
{\mathcal S}_j = \begin{cases}
\begin{pmatrix} 1 & s_j \\ 0 & 1\end{pmatrix}, & j = 0, \pm 2, \\[+1.em]
\begin{pmatrix} 1 & 0 \\ s_j & 1\end{pmatrix}, & j = \pm 1.
\end{cases}
\end{equation}

The Voros connection formula in (iii) is also well established
in the case of Schrodinger equation in \cite{KT05, Voros83}.
It enables us to describe these Stokes multipliers explicitly
via the Voros symbols~${\rm e}^{2\pi {\rm i} \nu/g_s}$ and ${\rm e}^{2\pi {\rm i} \rho/g_s}$.
A rough explanation is the following:
The Voros connection formula allows us to describe the analytic continuation
of the Borel resummed WKB solution by the term-wise analytic continuations on the spectral curve,
and the formula \eqref{eq:Voros-symbols} gives an explicit description
of those analytic continuations in terms of the Voros symbols
(see \cite[Section~5]{Iwaki19} for details.).
We will present the resulting Stokes multipliers in the next subsection.

\subsection{Delabaere--Dillinger--Pham formula and non-linear Stokes phenomenon}\label{subsection:DDP}

\begin{figure}[t]\centering
 \includegraphics[width=60mm]{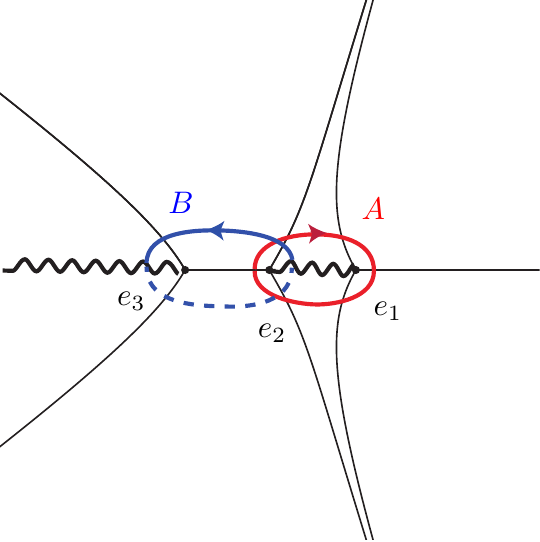}
 \caption{$A$-cycle and $B$-cycle.} \label{fig:cycles}
\end{figure}

Now, let us discuss the non-linear Stokes phenomenon for Painlev\'e~I.
We borrow the idea of Takei used in \cite{Takei1995, Takei1998}
where he relates the mutation of Stokes graphs (which we will explain shortly)
to non-linear Stokes phenomenon of Painlev\'e transcendents,
through the invariance of the Stokes multipliers of \eqref{eq:Lax-PI}.
This idea has been recently further developed in \cite{bssv,vv}.
We also note that our result is closely related to \cite{FIKN,Kapaev2004}
based on the Riemann--Hilbert method.

Figure \ref{fig:mutation} depicts Stokes graphs for $\nu = 1$
and several $t$'s which are close to $t_c = -5$. Let us label the three zeros of $\varphi(x)$ at $t = t_c$ as
\begin{equation}
e_1 \fallingdotseq 1.59075, \qquad
e_2 \fallingdotseq - 0.01288, \qquad
e_3 \fallingdotseq - 1.57157,
\end{equation}
and take the $A$ cycle (resp., the $B$ cycle) as the cycle encircling
the zeros $e_1$ and $e_2$ (resp., $e_2$ and $e_3$).
See Figure \ref{fig:cycles}.
Here the orientation of these cycle is given so that
\begin{equation}
\oint_A y {\rm d}x = 2 \int^{e_1}_{e_2} y {\rm d}x, \qquad
\oint_B y {\rm d}x = - 2 \int^{e_2}_{e_3} y {\rm d}x,
\end{equation}
where the branch of $y$ is chosen so that it has a positive imaginary part (resp., a negative real part)
on the segment $[e_2, e_1]$ (resp., $[e_3, e_2]$).
We may observe that a saddle connection appears at $t = t_c$,
where the $B$-period
\begin{equation}
\oint_B y {\rm d}x \fallingdotseq 5.87065
\end{equation}
of $y {\rm d}x$ has zero imaginary part.
As we have mentioned in Section~\ref{subsec:Stokes-graph}, we expect that
this induces singularities on the positive real axis on the Borel-plane,
and we will discuss the action of the Stokes automorphism below.

The saddle connection is resolved and we have saddle-free Stokes graphs
under a small variation of $t$ since the $B$-periods of $y dx$ are no longer real:
\begin{equation}% \label{eq:perturbed-B-periods}
\oint_B y {\rm d}x \fallingdotseq
\begin{cases}
5.83465 + 1.47447 {\rm i} & \text{at $t = t_c - {\rm i} \epsilon$}, \\
5.83465 - 1.47447 {\rm i} & \text{at $t = t_c + {\rm i} \epsilon$}.
\end{cases}
\end{equation}
Here we take $\epsilon = 1/2$.
We can observe that the topology of the Stokes graphs changes discontinuously
before and after the appearance of the saddle connection.
This is what we call the mutation of Stokes graph.

Since the Stokes graphs at $t = t_c \pm {\rm i} \epsilon$ are saddle-free,
we can use the recipe of \cite[Section~5]{Iwaki19} to compute the Stokes multipliers around $x= \infty$.
The resulting Stokes multipliers are
\begin{align}
&\text{At $t = t_c - {\rm i} \epsilon$} \colon\
\begin{cases}
s_{-2} = {\rm i} {\rm e}^{2\pi {\rm i} \nu/g_s}, \\
s_{-1} = {\rm i} \big({\rm e}^{- 2\pi {\rm i} \nu/g_s}
- {\rm e}^{-2\pi {\rm i} (\nu+\rho) /g_s} + {\rm e}^{- 2 \pi {\rm i} \rho/g_s}\big), \\
s_0 = {\rm i} {\rm e}^{2 \pi {\rm i} \rho/g_s}, \\
s_1 = {\rm i} {\rm e}^{- 2 \pi {\rm i} \rho/g_s} \big(1 - {\rm e}^{2 \pi {\rm i} \nu/g_s} \big), \\
s_2 = {\rm i} {\rm e}^{- 2 \pi {\rm i} \nu/g_s} \big(1 - {\rm e}^{2 \pi {\rm i} \rho / g_s}\big),
\end{cases}
\label{eq:Stokes-B-1}\\
&\text{At $t = t_c + {\rm i} \epsilon$} \colon\
\begin{cases}
{s}_{-2} = {\rm i} {\rm e}^{2\pi {\rm i} \nu/g_s} \big(1 - {\rm e}^{2 \pi {\rm i} \rho/g_s}\big),\\
{s}_{-1} = {\rm i} {\rm e}^{-2 \pi {\rm i} \rho/g_s} \big(1 - {\rm e}^{-2 \pi {\rm i} \nu/g_s}\big), \\
{s}_0 = {\rm i} {\rm e}^{2 \pi {\rm i} \rho/g_s}, \\
{s}_1= {\rm i} \big({\rm e}^{-2 \pi {\rm i} \rho/g_s}
- {\rm e}^{2 \pi {\rm i} (\nu - \rho)/g_s} + {\rm e}^{2 \pi {\rm i} \nu/g_s}\big), \\
{s}_2 = {\rm i} {\rm e}^{-2 \pi {\rm i} \nu/g_s}.
\end{cases}
\label{eq:Stokes-B-2}
\end{align}
Although Figure \ref{fig:mutation} is drawn for $\epsilon=1/2$,
the formulas \eqref{eq:Stokes-B-1} and \eqref{eq:Stokes-B-2}
are valid for any sufficiently small $\epsilon>0$
since the topological configuration of the Stokes graphs are same for those $\epsilon$.
We can check that the consistency conditions (cf.~\cite{SvdP})
\begin{equation}
1 + s_{j-1} s_j + {\rm i} s_j+2 = 0 ,\qquad s_j = s_{j+5}
\end{equation}
of the Stokes multipliers are satisfied for both cases.
We note that the result agrees with the computation of \cite[Section~7.4]{bssv}.
This is also consistent with \cite{Takei1998, vv} through the identification
of \cite[equations~(7.71)--(7.72)]{bssv}.

Thus, we have seen that the mutation of Stokes graphs induces
a discontinuous change of the expressions of the Stokes multipliers.
It is easy to observe that the Stokes data \eqref{eq:Stokes-B-2}
are obtained from \eqref{eq:Stokes-B-1} by the transformation
\begin{equation} \label{eq:DDP-B}
\big({\rm e}^{2 \pi {\rm i} \nu/g_s}, {\rm e}^{2 \pi {\rm i} \rho/g_s}\big) \mapsto
\big({\rm e}^{2 \pi {\rm i} \nu/g_s} \big(1 - {\rm e}^{2 \pi i \rho/g_s}\big), {\rm e}^{2 \pi {\rm i} \rho/g_s}\big).
\end{equation}
In fact, this is an example of a {cluster transformation}
(or a {Kontsevich--Soibelman transformation}).
Our observation is consistent with the results \cite{DDP93, DP99, GMN08, IN14, IN15}
where the mutation of Stokes graphs induces a cluster transformation
for the Voros symbols (the quantum period, the Fock--Goncharov coordinates).
\eqref{eq:DDP-B} is an example of
the Delabaere--Dillinger--Pham (DDP) formula.

Now we get to the crucial point.
Since $t$ is the isomonodromic time, these Stokes multipliers must be preserved under the variation of $t$.
Therefore, the above formulas suggest the following:
Let $(\nu^+, \rho^+)$ and $(\nu^-, \rho^-)$ be possibly different pairs of parameters,
and let ${\mathscr T}^{\pm}(t, \nu^\pm, \rho^\pm, g_s)$ be the Borel sum of
$\tau(t,\nu^\pm, \rho^\pm, g_s)$ defined at $t = t_c \pm {\rm i} \epsilon$.
If these $\tau$-functions correspond to a common solution of Painlev\'e I,
then they must be glued at $t = t_c$ so that the corresponding Stokes multipliers are identical.
Namely, we have
\begin{equation} \label{eq:Stokes-relation-tau}
{\mathscr T}_{P_{\rm I}}^{-}(t,\nu^-,\rho^-; g_s) =
{\rm e}^{\frac{1}{2 \pi {\rm i}} \operatorname{Li}_2({\rm e}^{2 \pi {\rm i} \rho^+/g_s})}
{\mathscr T}_{P_{\rm I}}^{+}(t, \nu^+, \rho^+; g_s)
\end{equation}
with
\begin{equation} \label{eq:monodromy-symplectomorphism}
( \nu^+, \rho^+ ) =
\left( \nu^- - \frac{g_s}{2 \pi {\rm i}} \log\big(1 - {\rm e}^{2\pi {\rm i} \rho^-/g_s}\big),\rho^- \right).
\end{equation}
Here, the prefactor $\exp\big(\frac{1}{2 \pi {\rm i}} \operatorname{Li}_2\big(e^{2 \pi {\rm i} \rho^+/g_s}\big)\big)$
on the right hand side is related to the generating function of the
monodromy symplectomorphism (which is the cluster transformation in this case).
The 2-form $2 \pi {\rm i} g_s^{-2} {\rm d} \nu \wedge {\rm d} \rho$ is a natural symplectic form
on the space of initial conditions for Painlev\'e I, and
\eqref{eq:monodromy-symplectomorphism} is a symplectic transform on the space.
It was proposed in \cite[equation~(3.17)]{LR-2016} that the (extended) $\tau$-function
for Painelv\'e I should be defined by taking the exponential of
the primitive of $- 2 \pi {\rm i} g_s^{-2} \nu {\rm d} \rho$ as a normalization factor
(see also \cite{BK-2019,boalch, CLT}).
The relation \eqref{eq:monodromy-symplectomorphism} implies
\begin{equation}
- \frac{2 \pi {\rm i}}{g_s^2} \left( \nu^{-} {\rm d}\rho^{-} - \nu^+ {\rm d}\rho^+ \right) =
{\rm d} \log {\rm e}^{\frac{1}{2 \pi {\rm i}} \operatorname{Li}_2({\rm e}^{2 \pi {\rm i} \rho^+/g_s})},
\end{equation}
and hence we have taken the prefactor in \eqref{eq:Stokes-relation-tau}.\footnote{We would like to thank I. Coman and F. Del Monte for helpful discussion on the prefactor.}

The formula \eqref{eq:Stokes-relation-tau} can be regarded as the connection formula which describes the
{non-linear Stokes phenomenon} for the Painlev\'e transcendents on the negative real axis in $t$-plane.
We may write the formula in terms of the Stokes automorphism as
\begin{equation}
\label{eq:Stokes-auto-tau}
{\mathfrak S} \tau(t,\nu,\rho ;g_s) =
{\rm e}^{\frac{1}{2 \pi {\rm i}} \operatorname{Li}_2(e^{2 \pi {\rm i} \rho/g_s})}
\tau\left( t, \nu - \frac{g_s}{2 \pi {\rm i}} \log\big(1 - e^{2\pi {\rm i} \rho/g_s}\big) , \rho; g_s \right).
\end{equation}
If we look at the zero Fourier mode (i.e., coefficient of $e^{2 \pi {\rm i} k \rho /g_s}$ with $k = 0$),
we have the all order instanton corrections of the partition function:
\begin{equation} \label{eq:Stokes-auto-ZTR}
{\mathfrak S} Z(t,\nu; g_s) = \sum_{n=0}^{\infty} Z^{(n)}(t,\nu ;g_s)
\end{equation}
with the terms of the form
\begin{equation}
 \label{eq:Stokes-terms}
Z^{(n)}(t,\nu ;g_s) = \mathfrak{Z}^{(n)}(t,\nu - n g_s; g_s) \,
Z(t,\nu - n g_s ; g_s).
\end{equation}
Here, $\mathfrak{Z}^{(n)}(t,\nu ;g_s)$ is a differential polynomial
of the free energy $F(t,\nu ;g_s)$ with respect to $\nu$.
The Seiberg--Witten relation \smash{$\partial_\nu F_0 = \oint_{B} y {\rm d}x$}
implies that $Z^{(n)}$ is a formal power series in $g_s$ with
an exponential factor \smash{${\rm e}^{- n \oint_B y {\rm d}x /g_s}$}.
Namely, $Z^{(n)}$ is an $n$-instanton amplitude.
It turns out that the first few terms
\begin{align}
Z^{(0)}(t,\nu; g_s) &
= Z(t,\nu; g_s),  \\
Z^{(1)}(t,\nu; g_s) &
= \left( 1 + \frac{g_s}{2\pi {\rm i}} \frac{\partial F}{\partial \nu}(t,\nu - g_s ; g_s) \right)
Z(t,\nu - g_s ; g_s)
\label{eq:1-instanton-TR}
\end{align}
of \eqref{eq:Stokes-auto-ZTR} precisely agree (up to a normalization factor)
with the multi-instanton results for the topological string obtained in \cite{gkkm,gm-multi}.
In the next section, we will show that the agreement occurs for all $n$ as well.
We may also observe that the first few terms of the 1-instanton part~\eqref{eq:1-instanton-TR}
are consistent with a known connection formula for Painlev\'e I (see, e.g., \cite{Kap, Kapaev2004}).
This supports our heuristic derivation of \eqref{eq:Stokes-auto-tau}.

Before ending the section, let us make a remark on a relation with the results of \cite{ bssv,Takei1998, vv}.
These works also derive a connection formula for solutions of Painlev\'e I
through the isomonodromy property.
Our result \eqref{eq:Stokes-auto-tau} describes the connection formula at the level of $\tau$-functions,
and the main difference is the appearance of the prefactor given by the dilogarithm function.
The factor disappears in the solution of Painlev\'e I due to the logarithmic derivative~\eqref{eq:q-and-tau}.
As we will see in the next section, we can relate the connection formula of Painlev\'e I
with the results of \cite{ gkkm,gm-multi} thanks to the prefactor.
This is our new observation.

\section[Resurgent structure and Stokes automorphisms in topological string theory]{Resurgent structure and Stokes automorphisms \\ in topological string theory}
\label{sec-HAEderivation}
In this section, we show that the main result from the previous section, \eqref{eq:Stokes-auto-tau}, is a consequence
of the conjectures of \cite{gkkm,gm-multi} on the resurgent structure of the topological string.

\subsection{Resurgent structure of the topological string}

In \cite{gkkm,gm-multi}, a general conjecture on the resurgent structure of the topological string
on arbitrary Calabi--Yau manifolds
was put forward. This conjecture is based on a trans-series solution of the HAE
of \cite{bcov}, as proposed in \cite{cesv2,cesv1}. For this reason, it applies to the free
energies obtained by doing topological
recursion on curves of genus $g \ge 1$, but it also applies to the free energies of compact Calabi--Yau
threefolds, since both are
perturbative solutions to the HAE. For simplicity, we will first
present the results in the one-modulus, local case originally studied in \cite{gm-multi}.
The generalization to the
multi-modulus, general case is straightforward and will be presented below.

The conjecture of \cite{gkkm,gm-multi} is as follows. First, it asserts that Borel singularities of the topological
string free energy are integral periods of the Calabi--Yau manifold, up to some overall normalization (in the local case,
this was already conjectured in~\cite{dmp-np}). In the
one-modulus, local case, this means that the
singularity ${\mathcal{A}}$ can be written as
\begin{equation}
\label{A-local}
{\mathcal{A}}= c \partial_\nu F_0 + d \nu+ d_0,
\end{equation}
where $c$, $d$, $d_0$ are integer numbers, times a normalization factor which depends
on the normalization of $g_s$ (see \cite{gkkm} for a detailed discussion of normalizations). We will assume that ${\mathcal{A}}$ is a~primitive vector
of the period lattice. Then, $\ell {\mathcal{A}}$, with
$\ell \in {\mathbb Z}_{>0}$ is also a Borel singularity, and we are interested in the structure of these ``multi-instanton'' singularities.
There are two different situations. When $c=0$, the resurgent
structure is of the Pasquetti--Schiappa form \cite{ps09}. This means the following. Let us define
\begin{equation}
\label{ps-instanton}
F^{(\ell)}_{\mathcal{A}}= \left( \frac{1}{\ell} \frac{{\mathcal{A}}}{g_s} + \frac{1}{\ell^2} \right) {\rm e}^{-\ell {\mathcal{A}}/g_s}.
\end{equation}
Then, the alien derivatives of the free energy are given by
\begin{equation}
\label{simple-alien}
\dot \Delta_{\ell {\mathcal{A}}} F= a F^{(\ell)}_{\mathcal{A}}, \qquad \ell \in {\mathbb Z}_{>0},
\end{equation}
where $a$ is a Stokes constant. When $c\not=0$, one defines a modified genus zero free energy $ \tilde F_0$ by the equation ${\mathcal{A}} = c \partial_\nu \tilde F_0$. We note that $ \tilde F_0$ differs from the original $F_0$ in a quadratic polynomial in $\nu$. The total free energy
appearing in the formulas for the
trans-series involves $\tilde F_0$, i.e., it is given by
\[
F(\nu ;g_s)= g_s^{-2} \tilde F_0(\nu) + \sum_{g \ge 1} g_s^{2g-2} F_g(\nu).
\]
Let us now define
\begin{equation}
\label{fell}
F^{(\ell)}(\nu; g_s) = \left( \frac{ \hbar F'(\nu - \ell \hbar;g_s)}{\ell} + \frac{1}{\ell^2}\right)
{\rm e}^{ F(\nu - \ell \hbar;g_s) - F(\nu; g_s)},
 \end{equation}
where we have introduced the rescaled coupling constant $\hbar= c g_s$. The prime in \eqref{fell} and other equations in this section denotes the derivative with reference to~$\nu$. Then, the alien derivatives of $F$ are given by
\begin{equation}
\label{alienfs}
\dot \Delta_{\ell {\mathcal{A}}} F= a F^{(\ell)}, \qquad \ell \in {\mathbb Z}_{>0}.
\end{equation}
These are the main conjectures of \cite{gkkm,gm-multi}. They recover and extend partial results along this direction in
\cite{cesv2,cesv1,cms}. We note that both in \eqref{alienfs} and \eqref{simple-alien} it is assumed that the Stokes constant
is independent of $\ell$.

From the formulas for the alien derivatives, one can compute the action of the Stokes automorphism,
through \'Ecalle's formula (see, e.g., \cite{msauzin})
\begin{gather}
\mathfrak{S}_{\mathcal C} = \exp \left( \sum_{\ell =1}^\infty {\mathcal C}^{\ell} \dot \Delta_{\ell {\mathcal{A}}} \right).
\end{gather}
We have introduced an additional formal parameter ${\mathcal C}$ to keep track of $\ell$. In view of the results of the
previous section, we should consider the action of the Stokes automorphism on the partition function, $Z$, and we have $\mathfrak{S}_{{\mathcal C}} (Z)= \exp \left( \mathfrak{S}_{{\mathcal C}} (F) \right)$. We have again two cases to consider. The simplest one is when $c=0$. In that case, the action of more than
one alien derivative vanishes, and we simply have
\begin{gather}
\mathfrak{S}_{{\mathcal C}} (Z)=\exp \left( a \sum_{\ell =1}^\infty {\mathcal C}^{\ell} F^{(\ell)}_{\mathcal{A}} \right)Z=\exp \left( a \sum_{\ell =1}^\infty {\mathcal C}^{\ell} \left( \frac{ 1}{\ell } \frac{{\mathcal{A}}}{g_s} + \frac{1}{\ell^2} \right) {\rm e}^{-\ell {\mathcal{A}}/g_s} \right)Z,
\end{gather}
which we can write as
\begin{equation}
\label{stokes-afr}
\mathfrak{S}_{{\mathcal C}} (Z)=
\exp \left( a \operatorname{Li}_2\big({\mathcal C} {\rm e}^{-{\mathcal{A}}/g_s}\big)- a \frac{{\mathcal{A}}}{g_s} \log\big(1-{\mathcal C} {\rm e}^{-{\mathcal{A}}/g_s}\big)\right)Z.
\end{equation}
This is the result obtained for the resolved conifold after using the Pasquetti--Schiappa form (see, e.g., \cite{astt}). We note that
the ingredients for our main formula (the dilogarithm and the logarithm) are already here,
and they follow from the Pasquetti--Schiappa form of the multi-instanton amplitudes, which is in turn
ultimately due to the universal behavior of the topological
string at the conifold point \cite{gv-conifold}.

The non-trivial case for the calculation of the Stokes automorphism occurs when $c\not=0$,
since one has to act with multiple alien
derivatives. Explicit calculations show that the result has the form
\begin{equation}
\label{stokesZ}
\mathfrak{S}_{{\mathcal C}} (Z)= \sum_{\ell \ge 0} {\mathcal C}^{\ell} \mathfrak{Z}^{(\ell)} (\nu- \ell \hbar) Z(\nu- \ell \hbar; g_s) ,
\end{equation}
where $\mathfrak{Z}^{(\ell)} (\nu)$ can be computed explicitly and $\mathfrak{Z}^{(0)}=1$. One finds, for the very first values,
\begin{gather}
\begin{split}
&\mathfrak{Z}^{(1)} = a \left(1+ \hbar F' \right), \\
& \mathfrak{Z}^{(2)}= \frac{a^2}{2} \left(1+2 \hbar F'+ (\hbar F')^2+ \hbar^2 F'' \right) + a \left( \frac{1}{4} + \frac{\hbar F'}{2} \right).
\end{split}
\end{gather}
This is in accord with \eqref{eq:Stokes-auto-ZTR}--\eqref{eq:1-instanton-TR}.
We will show in the next subsection that the structure \eqref{stokesZ} follows from the results of \cite{gm-multi}. The arguments
in Section \ref{painleve} suggest in addition the following explicit generating functional for the functions $\mathfrak{Z}^{(\ell)} (\nu)$:
\begin{equation}
\label{basic-for}
Z(\nu; g_s) \sum_{\ell \ge 0} {\mathcal C}^{\ell} \mathfrak{Z}^{(\ell)} (\nu)=
{\rm e}^{a \operatorname{Li}_2({\mathcal C})} Z\left(\nu - \hbar a \log(1-{\mathcal C}) ; g_s\right).
 \end{equation}
 Indeed, it follows from \eqref{stokesZ} and \eqref{basic-for} that
 \begin{equation}
\label{stokes-op}
\mathfrak{S}_{{\mathcal C}} (Z)= \exp\bigl\{ a \operatorname{Li}_2 \big({\mathcal C} {\rm e}^{-\hbar \partial} \big)-
 \hbar a \log \big(1- {\mathcal C} {\rm e}^{-\hbar \partial} \big) \partial \bigr\} Z,
\end{equation}
where $\partial$ is the derivative with reference to $\nu$. If we introduce now the discrete Fourier transform as in \eqref{eq:tau-PI},
\begin{gather}
\tau(\nu, \rho ;g_s) = \sum_{k \in {\mathbb Z}} {\rm e}^{2 \pi {\rm i} k \rho/g_s} Z(\nu+ k g_s; g_s),
\end{gather}
one easily finds
\begin{equation}
\label{stau-ts}
\mathfrak{S}_{{\mathcal C}} \tau(\nu, \rho;g_s)={\rm e}^{a \operatorname{Li}_2({\mathcal C} {\rm e}^{2 \pi {\rm i} c \rho / g_s} )}
 \tau \big(\nu -\hbar a \log\big(1- {\mathcal C} {\rm e}^{2 \pi {\rm i} c \rho/g_s}\big), \rho; g_s \big),
\end{equation}
where we have made a choice of normalizations in such a way that the coefficient $c$ is an integer. The formula
above has precisely the structure anticipated in \eqref{eq:Stokes-auto-tau}
(the two formulas agree after setting ${\mathcal C}=c=1$, $a = 1/(2 \pi {\rm i})$.)
In Section \ref{subsec-gens}, we will consider more general cases for the transformation of
the dual partition function and make contact with the results of \cite{AP}.

\subsection{A derivation of the formula for Stokes automorphisms}

We will now show that the formulas \eqref{basic-for} and \eqref{stokes-op} follow from the conjecture on the alien
derivatives of the free energy, \eqref{alienfs}. To do this, we will rely on various results of \cite{gkkm,gm-multi}, which we
summarize very briefly here. We refer to those papers for more details.

In the framework of the HAE,
the perturbative free energies $F_g$ are non-holomorphic but global functions on the moduli space.
More precisely, they are polynomials in a non-holomorphic propagator $S$, whose coefficients are
functions of a complex coordinate $z$ on the moduli space (not necessarily flat).
The conventional free energies are holomorphic but can be defined in different frames,
which are determined by a choice of $A$ and $B$ periods. The holomorphic free energies in different frames
are obtained by considering the non-holomorphic free energies, and taking different
holomorphic limits of the propagator. We
define the ${\mathcal{A}}$-frame as the frame in which ${\mathcal{A}}$ is the $A$-period, and the holomorphic
propagator appropriate for the ${\mathcal{A}}$-frame will be denoted by $S_{\mathcal{A}}$. The
boundary conditions to solve the HAE are obtained by evaluating the
holomorphic free energies in different frames and imposing particular behaviors at special points in moduli space, in particular
the universal behaviour at the conifold point \cite{gv-conifold} and the resulting gap condition \cite{hk06}.

It was pointed out in \cite{cesv2,cesv1,cms} that the resurgent structure of the
topological string free energy can be obtained
by considering trans-series solutions to the HAE of \cite{bcov}. The solutions
corresponding to the $\ell$-th instanton sector are formal power series
in $g_s$, whose coefficients are also polynomials in $S$ with $z$-dependent coefficients, and they
also involve ${\mathcal{A}}$ and their derivatives (the second derivative of ${\mathcal{A}}$ can be however
re-expressed in terms of $S_{\mathcal{A}}$). The $\ell$-instanton amplitude involves of course an
exponential prefactor of the form ${\rm e}^{-\ell {\mathcal{A}}/g_s}$. Explicit trans-series solutions were obtained in
\cite{gkkm,gm-multi} by using an
operator formalism first suggested in \cite{codesido-thesis,coms}. The main operator in
this formalism is\footnote{For the factors of $g_s$, we follow the conventions in \cite{gkkm}.}
\begin{gather}
\mathsf{D} = g_s \partial_z {\mathcal{A}} (S- S_{\mathcal{A}}) \partial_z.
\end{gather}
When evaluated in the holomorphic limit, this operator becomes $\hbar \partial_\nu$, i.e., a derivative with reference to the flat coordinate $\nu$.

We are now ready to prove the formula \eqref{basic-for}. The Stokes
automorphism, acting on $Z$, produces a formal sum of
multi-instanton sectors which has to solve the HAE for the partition function,
and in addition it has to satisfy the following boundary condition: when
evaluated at the ${\mathcal{A}}$-frame, it is equal to \eqref{stokes-afr}. This determines its form uniquely.
A general $n$-th instanton solution to the HAE for the partition function
was determined in \cite{gm-multi} and has the structure:
\begin{equation}
\label{gen-sol}
 Z^{(n)}= \mathfrak{A}_n \, {\rm e}^{-\Phi_n} Z,
\end{equation}
where
\begin{gather}
\Phi_n=\frac{1}{\mathsf{D}} \big(1- {\rm e}^{-n \mathsf{D}} \big) G,
\qquad G= \frac{1}{g_s} {\mathcal{A}}+ \mathsf{D} \big(F- g_s^{-2} F_0\big).
\end{gather}
Let us note that, in the holomorphic limit, $\Phi_n \rightarrow F(\nu ;g_s)- F(\nu-n \hbar ; g_s)$. The prefactor $\mathfrak{A}_n$ is determined as follows. Let $X_n ={\rm e}^{-n \mathsf{D}} G$. Then, the $\mathfrak{A}_n$ are arbitrary linear combinations of the objects $\mathfrak{w}_\ell$, defined by
\begin{gather}
 \mathfrak{w}_\ell= \sum_{\boldsymbol{k}, \, d(\boldsymbol{k})=\ell}
 C_{\boldsymbol{k}} \mathfrak{X}_{\boldsymbol{k}}.
 \end{gather}
In this formula, $\boldsymbol{k}=(k_1, k_2, \ldots)$ is a vector of non-negative entries, $d(\boldsymbol{k})$ is given by $d(\boldsymbol{k})=\sum_j j k_j$,
 the coefficients $C_{\boldsymbol{k}}$ are of the form,
\begin{gather}
 C_{\boldsymbol{k}}= \frac{\ell!}{\prod_{j \ge 1} k_j! (j!)^{k_j}},
 \end{gather}
and $\mathfrak{X}_{\boldsymbol{k}} =X_n^{k_1} (\mathsf{D} X_n)^{k_2} \big(\mathsf{D}^2 X_n\big)^{k_3} \cdots$. We would like to emphasize that this structure is determined by requiring that \eqref{gen-sol} is a~solution to the HAE (we recall that the HAE for the partition function is linear, so we can solve it for each instanton sector separately). Let us also
note that all the $X$'s appearing in $\mathfrak{A}_n$ are shifted, i.e., they are acted upon by
the automorphism ${\rm e}^{-n \mathsf{D}}$, so it is convenient to introduce the ``unshifted'' prefactor $\mathfrak{B}_n$ defined by
\begin{equation}\label{ABfrak}
\mathfrak{A}_n= {\rm e}^{-n \mathsf{D}} \mathfrak{B}_n.
\end{equation}
In $\mathfrak{B}_n$, $X_n$ is replaced by $X=G$. The precise linear combination of
$\mathfrak{w}_\ell$ appearing in $\mathfrak{A}_n$ is uniquely determined by the boundary condition, i.e., by
its form in the ${\mathcal{A}}$-frame. As we will see in a moment, when evaluated in the ${\mathcal{A}}$-frame, the holomorphic limit of
$Z^{(n)}$ is of the form
\begin{equation}
\label{bound-form}
Z^{(n)}_{\mathcal{A}} = \left\{ \sum_{k} c_{n,k} \left( \frac{ {\mathcal{A}}}{g_s} \right)^k \right\} {\rm e}^{-n {\mathcal{A}}/g_s} Z_{\mathcal{A}},
\end{equation}
where the prefactor is an arbitrary polynomial in ${\mathcal{A}}/g_s$. Then, $\mathfrak{B}_n= \sum_k c_{n,k} \mathfrak{w}_k$.

Let us now apply these results to the calculation of the Stokes automorphism. As we explained before,
the Stokes automorphism produces the holomorphic limit of a formal linear combination of solutions of the form
\eqref{gen-sol}. Therefore, we must have
\begin{gather}
\mathfrak{S}_{{\mathcal C}}(Z) = \sum_{\ell \ge 0} {\mathcal C}^\ell \mathfrak{A}_{\ell} Z(\nu- \hbar \ell;g_s).
\end{gather}
(To lighten the notation, we are using the same symbols for the non-holomorphic quantities
appearing in the HAE, and for their holomorphic limits. Hopefully, which one is being used at a given moment is
clear from the context.) This is precisely the structure of \eqref{stokesZ}, which follows from the general
results for multi-instantons. We deduce that $\mathfrak{A}_{\ell} = \mathfrak{Z}^{(\ell)} (\nu- \hbar \ell )$. Both sides involve functions whose argument is shifted by $-\hbar \ell$. In terms of the unshifted prefactors
introduced in \eqref{ABfrak}, we have
\begin{equation}
\mathfrak{B}_\ell = \mathfrak{Z}^{(\ell)} (\nu).\label{BZ}
\end{equation}
The boundary condition obtained from \eqref{stokes-afr} is
\begin{align}
\sum_{n \ge 0} {\mathcal C}^n \mathfrak{B}_{n, {\mathcal{A}}}={}&\exp \left( a \operatorname{Li}_2({\mathcal C})- a \frac{{\mathcal{A}}}{g_s} \log(1-{\mathcal C})\right)\nonumber\\
={}& {\rm e}^{a \operatorname{Li}_2 ({\mathcal C})}
 \sum_{k \ge 0} \frac{1}{k!} \left( - a \log(1-{\mathcal C}) \right)^k \left( \frac{{\mathcal{A}}}{g_s} \right)^k,
\end{align}
which is indeed of the form \eqref{bound-form}. According to what we explained above,
we can already write the general
solution to the HAE, by simply replacing $({\mathcal{A}}/g_s)^k$ by $\mathfrak{w}_k$:
\begin{gather}
\sum_{n \ge 0} {\mathcal C}^n \mathfrak{Z}^{(n)} =
{\rm e}^{a \operatorname{Li}_2 ({\mathcal C})} \sum_{k \ge 0} \frac{1}{k!} \left( - a \log(1-{\mathcal C}) \right)^k \mathfrak{w}_k,
\end{gather}
where we have already used \eqref{BZ}. It was proven in \cite{gm-multi} that
\begin{gather}
\Xi(\xi)= \sum_{\ell \ge 0} \frac{\xi^\ell}{\ell!} \mathfrak{w}_\ell= \exp \left( \sum_{j=1}^\infty \frac{\xi^j}{j!} \mathsf{D}^{j-1}
X \right),
\end{gather}
where $\xi$ is an arbitrary complex parameter. We conclude that
\begin{gather}
\sum_{n \ge 0} {\mathcal C}^n \mathfrak{Z}^{(n)} ={\rm e}^{a \operatorname{Li}_2({\mathcal C})}
\exp \left( \frac{1}{\mathsf{D}} \big( {\rm e}^{-a \log(1-{\mathcal C}) \mathsf{D}}- 1 \big) X \right).
\end{gather}
In the holomorphic limit, we have that $X \rightarrow \hbar \partial_\nu F$, where $F$ is the total free energy, and we get in the end
\begin{gather}
\sum_{n \ge 0} {\mathcal C}^n \mathfrak{Z}^{(n)} = \exp \left( a \operatorname{Li}_2({\mathcal C}) + F(\nu-a \hbar \log(1-{\mathcal C});g_s) - F(\nu;g_s) \right).
\end{gather}
This is precisely \eqref{basic-for}.

\subsection{Generalizations}\label{subsec-gens}

The above results concern the one-modulus, local case. However, the generalization to
arbitrary CY threefolds is straightforward,
by using the results of \cite{gkkm} (to which we refer for further details). We will now write in some
detail the more general formula for the Stokes automorphism. In the case of an arbitrary CY, the genus
$g$ free energies depend on
the ``big moduli space'' flat coordinates $X^I$ of the CY, where $I=0,1, \dots, n$. The
Borel singularities or instanton actions are
again integral periods, given by linear combinations,
\begin{equation}\label{A-global}
\kappa^{-1} {\mathcal{A}}=c^I \frac{\partial F_0}{\partial X^I}+ d_I X^I.
\end{equation}
We note that, in the local case, $X^0=1$, and $\partial F_0/ \partial X^0$ does not appear, so in the one-modulus case this expression reduces to \eqref{A-local}. In \eqref{A-global}, we have introduced explicitly the
normalization factor $\kappa$ relating the action to the integral periods.
If all $c^I=0$, the multi-instanton amplitudes are again of the Pasquetti--Schiappa form \eqref{ps-instanton} and the
Stokes automorphism is given by \eqref{stokes-afr}. When not all $c^I$ vanish, one defines a new genus zero free energy by \smash{${\mathcal{A}}=\kappa c^I \frac{\partial \tilde F_0}{\partial X^I}$}, as in the local case. It can be written as
\begin{gather}
\tilde F_0 \big(X^I\big)= F_0 \big(X^I\big) + \frac{1}{2} a_{IJ} X^I X^J, \qquad a_{IJ} c^I= d_J.
\end{gather}
Of course, the final formulas will not depend on $a_{IJ}$, but only on $c^I$, $d_J$.
As shown in \cite{gkkm}, one has to define a new genus one free energy
\begin{gather}
\tilde F_1= F_1-\left( \frac{\chi}{24} -1 \right) \log X^0.
\end{gather}
Such a redefinition has appeared before, e.g., in\cite[equation~(2.77)]{APP}. The total free energy
relevant for the multi-instanton amplitudes will be denoted by $\tilde F \big(X^I;g_s\big)$, and is given by
\begin{align}
 \widetilde F\big(X^I; g_s\big)&= g_s^{-2} \tilde F_0\big(X^I\big) + \tilde F_1\big(X^I\big)+ \sum_{g \ge 2} g_s^{2g-2} F_g\big(X^I\big)\nonumber\\
 &= \frac{1}{2g_s^2} a_{IJ} X^I X^J + F \big(X^I;g_s\big).\label{FFtilde}
\end{align}
Then, one has the following generalization of \eqref{stokes-op},
\begin{equation}
\label{mfrakmulti}
\mathfrak{S}_{{\mathcal C}} (\tilde Z)= \exp\bigl\{ a \operatorname{Li}_2 \bigl({\mathcal C} {\rm e}^{-\kappa g_s c^I\partial_I } \bigr)-
a \kappa g_s \log \bigl(1- {\mathcal C} {\rm e}^{-\kappa g_s c^I \partial_I} \bigr) c^I \partial_I \bigr\} \tilde Z,
 \end{equation}
 where we have denoted $\tilde Z= {\rm e}^{\tilde F}$, and $\partial_I= \frac{\partial}{\partial X^I}$.

As we have explained before, the action of the Stokes automorphism has
a simpler form when it acts on an appropriate dual partition function. We could obtain
a direct generalization of~\eqref{stau-ts} involving
the redefined partition function~$\tilde Z$. However, in order to make contact with
the results of~\cite{AP}, it is convenient to
consider the dual partition function to the original~$Z$. This means that in~\eqref{mfrakmulti} we have to treat separately the
quadratic term in $X^I$ appearing in the second line of~\eqref{FFtilde}. If we denote
\begin{gather}
Y^I= X^I- a \kappa g_s c^I \log (1- {\mathcal C} ),
\end{gather}
we find
\begin{align}
\tilde Z \big(Y^I; g_s \big)={}&\exp\left\{ \frac{\kappa^2}{2} a^2 d^I c_I \log^2 (1- {\mathcal{C}}) - g_s^{-1} a \kappa \log (1- {\mathcal{C}}) d_I X^I +\frac{1}{2 g_s^2} a_{IJ} X^I X^J \right\}\nonumber\\
&\times Z \big( Y^I ;g_s \big).
\end{align}
We also note that, for $n \in {\mathbb Z}$,
\begin{gather}
{\rm e}^{-n \kappa g_s c^I \partial_I} \exp \left\{ \frac{1}{2 g_s^2} a_{IJ} X^I X^J\right\}=
 \exp \left\{ -g_s^{-1} \kappa d_I X^I n +\frac{n^2}{2} \kappa^2 c^I d_I \right\}\nonumber\\
 \hphantom{{\rm e}^{-n \kappa g_s c^I \partial_I} \exp \left\{ \frac{1}{2 g_s^2} a_{IJ} X^I X^J\right\}=}{}
 \times \exp \left\{ \frac{1}{2 g_s^2} a_{IJ} X^I X^J\right\}.
\end{gather}
We have to be more concrete about the normalization factor $\kappa$. It was found in \cite{gkkm} that,
with the canonical normalization of $g_s$, one has $\kappa^2= - 2 \pi {\rm i}$, and this means that
\begin{gather}
 {\rm e}^{\frac{n^2}{2} \kappa^2 c^I d_I }= {\rm e}^{ -\pi {\rm i} n c^I d_I },
 \end{gather}
 since $n, d_I, c^I \in {\mathbb Z}$. We will now put together $c^I$, $d_I$ in a symplectic vector $\gamma= \big(c^I, d_I\big)$. Let us introduce
\begin{gather}
 \boldsymbol{X}_\gamma=\sigma(\gamma) \exp \big[-\kappa g_s^{-1} \big( d_I X^I - \rho_I c^I \big) \big],
 \end{gather}
 where $\rho_I$, $I=0,1, \dots, n$, are additional variables, and $\sigma(\gamma)= (-1)^{d_I c^I}$. Then, one finds that~\eqref{mfrakmulti} is equivalent to
\begin{align}%\label{big-trans}
&\mathfrak{S}_{{\mathcal C}} \left(\sum_{\boldsymbol{\ell} \in {\mathbb Z}^n} {\rm e}^{ \kappa \rho_I \ell^I/g_s} Z \big(X^I+ \kappa \ell^I g_s ;g_s \big) \right)\nonumber\\
&\qquad={\rm e}^{a \operatorname{Li}_2({\mathcal C} \boldsymbol{X}_\gamma) - \pi {\rm i} a^2 d^I c_I \log^2 (1- {\mathcal C} \boldsymbol{X}_\gamma) } \sum_{\boldsymbol{\ell} \in {\mathbb Z}^n} {\rm e}^{- a g_s^{-1} \kappa \log (1- {\mathcal C} \boldsymbol{X}_\gamma) d_I (X^I + \kappa g_s \ell^I )}\nonumber\\
 &\phantom{\qquad=}{}\times Z \left(X^I + \kappa g_s \ell^I - a g_s \kappa c^I \log\left(1- {\mathcal C} \boldsymbol{X}_\gamma \right);g_s \right){\rm e}^{ \kappa \rho_I \ell^I/g_s}.
 \end{align}
The appropriate definition of the dual partition function
in this general case is~\cite{AP}:
\begin{gather}
\tau\big(X^I, \rho_I;g_s\big)= {\rm e}^{ \frac{1}{2 g_s^2} X^I \rho_I}
\sum_{\boldsymbol{\ell} \in {\mathbb Z}^n}
{\rm e}^{ \kappa \rho_I \ell^I/g_s}
Z \big(X^I+ \kappa \ell^I g_s ;g_s \big),
\end{gather}
and the action of the Stokes automorphism is
\begin{align}
\mathfrak{S}
\tau\big(X^I, \rho_I;g_s\big)={}& \exp( a L_{\sigma(\gamma) } ( \boldsymbol{X}_\gamma ) )\nonumber\\
&\times \tau
\big( X^I- a g_s \kappa c^I \log(1- \boldsymbol{X}_\gamma ), \rho_I-a g_s \kappa d_I \log(1- \boldsymbol{X}_\gamma ) ;g_s \big),\label{gen-stokes-AP}
\end{align}
where we have put ${\mathcal C}=1$, and $L_{\epsilon}(z)$ is the twisted Rogers dilogarithm, as in \cite{AP}:
\begin{gather}
L_{\epsilon}(z)= \operatorname{Li}_2(z)+ \frac{1}{2} \log\big( \epsilon^{-1} z \big) \log(1-z).
\end{gather}
It is easy to see that \eqref{gen-stokes-AP} agrees precisely with the wall-crossing formula (1.9) in \cite{AP},
where their variables $\xi^I$, $\tilde \xi_I$ are related to ours by $\big(X^I, \rho_I\big)= - \kappa g_s \big(\xi^I, \tilde \xi_I\big)$.\footnote{\cite{AP} also
give a wall-crossing formula for $Z$, in terms of an integral transform, which is equivalent to \eqref{mfrakmulti}. We would like
to thank Boris Pioline for many discussions on the relation between the approach of \cite{APP,AP} and the one presented in this paper.}
In addition, the agreement between the formulas requires the identification \eqref{aomega-intro}.
One advantage of \eqref{gen-stokes-AP} is that, when~${c^I=0}$, one recovers as well the
transformation formula \eqref{stokes-afr} (this is easily seen by looking, e.g., at the mode with $\ell^I=0$).

There is another generalization of the formula \eqref{basic-for} that one could consider. So far we have only included forward alien derivatives,
and correspondingly purely instanton sectors. We can also consider alien derivatives in both the
negative and the positive directions, which lead to amplitudes with both instantons and ``negative instantons''.
Let us define
\begin{equation}\label{anti-sign}
 F^{(0|\ell)} (\nu; g_s)=-F^{(\ell)}(\nu;-g_s), \qquad F^{(0|\ell)}_{\mathcal{A}} (\nu;g_s)= -F^{(\ell)}_{\mathcal{A}} (\nu;-g_s).
 \end{equation}
The basic alien derivative in the negative direction is simply,
\begin{equation}\label{neg-alien}
\dot \Delta_{-\ell {\mathcal{A}}} F= a F^{(0|\ell)} \quad \text{or} \quad a F^{(0|\ell)}_{\mathcal{A}},
\end{equation}
depending on whether $c\not=0$ or $c=0$. We can then consider the ``mixed''
Stokes automorphism:
\begin{gather}
\mathfrak{S}_{{\mathcal C}_1, {\mathcal C}_2}= \exp \Bigg( \sum_{\ell \ge 1} {\mathcal C}_1^\ell \dot \Delta_{\ell {\mathcal{A}}} + {\mathcal C}_2^\ell \dot \Delta_{-\ell {\mathcal{A}}} \Bigg).
\end{gather}
Acting on $Z$, it has the structure
\begin{gather}
\mathfrak{S}_{{\mathcal C}_1, {\mathcal C}_2} (Z)= \sum_{\ell_1, \ell_2 \ge 0} {\mathcal C}_1^{\ell_1} {\mathcal C}_2^{\ell_2} \mathfrak{Z}^{(\ell_1|\ell_2)} (\nu- (\ell_1- \ell_2) \hbar) Z (\nu- (\ell_1- \ell_2) \hbar ),
\end{gather}
where $\mathfrak{Z}^{(0|0)}=1$. In this case, the boundary condition follows from \eqref{neg-alien}, \eqref{anti-sign} and
\eqref{ps-instanton}. It is given by
\begin{align}
&\exp \left\{ a \sum_{\ell =1}^\infty \big( {\mathcal C}_1^{\ell} F^{(\ell)}_{\mathcal{A}}+{\mathcal C}_2^{\ell} F^{(0 |\ell)}_{\mathcal{A}} \big) \right\}\nonumber\\
&\qquad=\exp \left\{ a \left( \operatorname{Li}_2 ({\mathcal C}_1) - \operatorname{Li}_2 ({\mathcal C}_2) \right) - a \frac{ {\mathcal{A}}}{g_s} \left( \log(1- {\mathcal C}_1)+ \log(1- {\mathcal C}_2)\right) \right\}.
\end{align}
By using the results in \cite{gm-multi}, we can generalize \eqref{basic-for} to
\begin{align}
&Z( \nu; g_s) \sum_{\ell_1, \ell_2 \ge 0} {\mathcal C}_1^{\ell_1} {\mathcal C}_2^{\ell_2} \mathfrak{Z}^{(\ell_1| \ell_2)} (\nu) \nonumber\\
& \qquad= {\rm e}^{ a \left(\operatorname{Li}_2({\mathcal C}_1)-\operatorname{Li}_2({\mathcal C}_2)\right) }
Z \left( \nu- a \hbar \log (1- {\mathcal C}_1)- a \hbar \log(1-{\mathcal C}_2); g_s \right).
\end{align}
It is easy to write this formula in the form \eqref{mfrakmulti} or \eqref{gen-stokes-AP}.

\appendix
\section[Definition of correlators and free energy in topological recursion]{Definition of correlators and free energy\\ in topological recursion}
\label{appendix:TR}

To apply the topological recursion, we regard \eqref{eq:spcurve}
as a family of spectral curves in the sense of~\cite{EO07}
(i.e., a data consisting of a compact Riemann surface $C$
and a pair $(x, y)$ of meromorphic functions on it),
through the Weierstrass parametrization:
\begin{gather}%\label{eq:par-rep}
C = {\mathbb C}/\Lambda, \qquad
x(z) = \wp(z), \qquad y(z) = \wp'(z).
\end{gather}
Here $\wp(z) = \wp(z; g_2, g_3)$ is the Weierstrass $\wp$-function
with $g_2=-2t$ and $g_3 = - u$, which is doubly-periodic
with periods $\omega_A$ and $\omega_B$ (we omit the $t$ and $\nu$ dependence for simplicity).
$\Lambda = {\mathbb Z} \omega_A + {\mathbb Z} \omega_B$ is the lattice
generated by the periods of the elliptic curve \eqref{eq:spcurve}.

Let $z_o \in {\mathbb C}$ be a generic point,
and $\Omega$ be the quadrilateral with
$z_o$, $z_o+\omega_A$, $z_o+\omega_B$ and~${z_o+\omega_A+\omega_B}$
on its vertices; that is, a fundamental domain of ${\mathbb C}/\Lambda$.
The ramification points (i.e., zeros of ${\rm d} x$) on $\Omega$
are given by the half-periods
$r_1 \equiv \omega_A/2$, $r_2 \equiv \omega_B/2$ and $r_3 \equiv (\omega_A+\omega_B)/2$
modulo $\Lambda$. These points correspond to the branch points
$e_i = x(r_i)$ ($i=1,2,3$) of the elliptic curve which are defined by
$4x^3+2tx+u = 4(x-e_1)(x-e_2)(x-e_3)$.
The covering involution~${y \mapsto -y}$ is realized by
$z \mapsto \sigma(z) \equiv -z$ mod $\Lambda$.

To run the topological recursion, we also need
the {Bergman bidifferential} normalized along the chosen A-cycle.
For our spectral curve, it is given by
\begin{gather}% \label{eq:Bergman-kernel}
B(z_1, z_2) = \left( \wp(z_1 - z_2)
+ \frac{\eta_A}{\omega_A} \right) {\rm d}z_{1} {\rm d}z_2,
\end{gather}
where \[\eta_A = - \oint_A \frac{x {\rm d}x}{y}.\]
Then, the
{topological recursion}
recursively constructs a doubly-indexed sequence of
meromorphic multi-differentials $\omega_{g,n}(z_1, \dots, z_n)$
($g \ge 0$, $n \ge 1$) on the spectral curve, called {correlators}, as follows:
\begin{gather}
\omega_{0,1}(z_{1}) = y(z_{1}) {\rm d}x(z_{1}), \qquad
\omega_{0,2}(z_{1},z_{2}) =
B(z_1, z_2),
\end{gather}
and for $2g-2+n \ge 1$, we define
\begin{gather}%\label{eq:top-rec}
\omega_{g,n}(z_{1},\dots,z_{n}) =
\sum_{j=1}^{3}
\Res_{z=r_j} K(z_1,z) R_{g,n}(z, z_2, \dots, z_n),
\end{gather}
where
\begin{align}
R_{g,n}(z, z_2, \dots, z_n) ={}&
\omega_{g-1,n+1}(z, \sigma(z),z_2, \dots, z_n) \nonumber\\
&+
\sum'_{\substack{g_{1}+g_{2}=g \\ I \sqcup J = \{2,\dots,n \}}}
\omega_{g_{1}, |I|+1}(z,z_{I}) \omega_{g_{2}, |J|+1}( \sigma(z), z_{J}).\label{eq:R-gn}
\end{align}
Here, the {recursion kernel} $K(z_1,z)$ is given by
\begin{gather} %\label{eq:recursion-kernel}
K(z_1,z) =
\frac{1}{(y(z) - y( \sigma(z))) {\rm d}x(z)}
\int^{w=z}_{w=0} \omega_{0,2}(z_{1}, w).
\end{gather}
We use the convention for a tuple of variables as
$z_{I} = (z_{i_1}, \dots, z_{i_{k}})$ if
$I = \{i_1, \dots, i_k \}$,
and the prime in the right-hand side of \eqref{eq:R-gn}
means that only indices satisfying $(g_i, I_i) \ne (0, \varnothing)$ are taken
(i.e., $\omega_{0,1}$ does not appear) in the summation.

Here we also recall the definition of the genus $g$ free energy
$F_g = F_g(t,\nu)$ introduced in \cite{EO07}.
The {genus $0$ free energy} $F_0$ is defined in
\cite[Section~4.2.2]{EO07}. In our case, it is given by
\begin{gather} %\label{eq:expression-F0}
F_0 = \frac{t u}{5} + \frac{\nu}{2} \oint_B y {\rm d}x.
\end{gather}
The {genus $1$ free energy} $F_1$ is also defined in \cite[Section~4.2.3]{EO07}
up to a multiplicative constant. We employ
\begin{gather} %\label{eq:expression-F1}
F_1 = - \frac{1}{12} \log \big(\omega_A^6 {\mathcal D} \big)
\end{gather}
as the definition.
Here
\begin{gather}
{\mathcal D} = -8t^3 -27 u(t,\nu)^2
\end{gather}
is the discriminant of \eqref{eq:spcurve}.
Finally, we define the {genus $g$ free energy} $F_{g}$ for $g \ge 2$ by
\begin{gather} % \label{eq:Fg}
F_{g} = \frac{1}{2-2g}
\sum_{j=1}^{3} \Res_{z=r_j} \Phi(z) \omega_{g,1}(z),
\end{gather}
where $\Phi$ is any primitive of $\omega_{0,1}$.
See \cite{EO07} for properties of $\omega_{g,n}$ and $F_g$.

\subsection*{Acknowledgements}
We would like to thank Ioana Coman, Fabrizio del Monte, Alba Grassi, Paolo Gregori, Jie Gu,
Shinobu Hosono, Omar Kidwai, Oleg Lisovyy, Pietro Longhi, Kento Osuga, Boris Pioline, Ricardo Schiappa,
Masa-Hiko Saito, Maximilian Schwick, Atsushi Takahashi, Yoshitugu Takei and Joerg Teschner
for useful conversations.
We would also like to thank to Maxim Kontsevich and Yan Soibelman
for organizing the IHES school ``Wall-crossing structures, analyticity and resurgence'',
which made possible this collaboration.
The work of K.I has been supported by the JSPS KAKENHI Grand Numbers
20K14323, 21K18576, 21H04994, 22H00094, 23K17654. The work of M.M. has been
supported in part by the ERC-SyG project ``Recursive and Exact New
Quantum Theory'' (ReNewQuantum), which received funding from the
European Research Council (ERC) under the European Union's Horizon
2020 research and innovation program, grant agreement No. 810573.
Finally, we would like to thank the anonymous referees
for their helpful comments that improved the quality of the paper.

\pdfbookmark[1]{References}{ref}
\LastPageEnding

\end{document}